\documentstyle[aps,pre,eqsecnum]{revtex}
\begin{document}
\draft
\title{Dynamic Fluctuation Phenomena in Double Membrane Films}

\author{E. I. Kats*\dag\dag, V. V. Lebedev*\dag, and S. V. Malinin*}

\address{*L. D. Landau Institute for Theoretical Physics, RAS \\
117940, Kosygina 2, Moscow, Russia; \\
\dag Department of Physics, Weizmann Institute of Science,
76100, Rehovot, Israel; \\
\dag\dag Max-Planck Institute for Physics of Complex Systems,
D-01187 Dresden, Germany.}
\date{\today}

\maketitle

\begin{abstract}
Dynamics of double membrane films is investigated in the long-wavelength
limit $qh\ll1$ ($q$ is the wave vector and $h$ is the thickness of the film)
including the over damped squeezing mode. We demonstrate that thermal
fluctuations essentially modify the character of the mode due to its
nonlinear coupling to the transversal shear hydrodynamic mode. The
renormalization can be analysed under condition $g\ll1$ (where
$g\sim T/\kappa$, $T$ is the temperature and $\kappa$ is the bending module).
The corresponding Green function acquires as a function of the frequency
$\omega$ a cut along the imaginary semi-axis. At $qh>\sqrt{g}$ the effective
length of the cut is $\sim Tq^3/\eta$ (where $\eta$ is the shear viscosity of
the liquid). At $qh<\sqrt{g}$ fluctuations lead to increasing the attenuation
of the squeezing mode: it is larger than the `bare' value by the factor
$1/\sqrt{g}$.
\end{abstract}


\section{Introduction}

The most distinctive property of amphiphilic molecules is their ability to
spontaneously self-assemble into aggregates of various shapes. Typically the
molecules spontaneously self-assemble into membranes which are bilayers of a
thickness of the order of a molecular length. Different lyotropic structures
constituted of the membranes have generated considerable current interest
(see the books \cite{ML87,SC87,NP89} and the reviews \cite{BP84,PA91,PO92}).
Films composed of two bilayer membranes sandwiching a thin layer of a liquid
are widely spread in the lyotropic systems. They play also an essential role
for various biological processes (one can note the so-called flickering
phenomena in erythrocytes or red blood cells). In the paper we will examine
dynamic properties of such double membrane films.

The main peculiarity of a membrane is its negligible surface tension. Indeed,
the membrane is immersed into a liquid and consequently its area can vary.
Zero surface tension is the equilibrium condition with respect to the
variations. In the situation shape fluctuations of the membrane are
determined by the bending elasticity, the corresponding energy is
\cite{CA70,HE75}
\begin{eqnarray}
{\cal H}_{\rm curv}= \frac{\kappa}{2}\int dA \,
\left(\frac{1}{R_1}+\frac{1}{R_2} \right)^2 \,,
\label{a1} \end{eqnarray}
where the integral is taken over the membrane which is considered as a
two-dimensional object, $R_1, R_2$ are its local curvature radii, and
$\kappa$ is the bending rigidity module. Corrugations of the membrane induced
by the thermal noise lead to loosing the orientation correlation of the
membrane pieces at separations larger than the so-called persistent length
$\xi_p$ \cite{GT82} which can be estimated as
$$ \xi_p\sim a \exp(2\pi\kappa/T) \,, $$
where $T$ is the temperature and $a$ is the thickness of the membrane. The
shape fluctuations of the membrane lead to the logarithmic renormalization of
the bending module $\kappa $, examined first by Helfrich \cite{HE85} and
later by F\"orster \cite{FO86}, the correct renormalization-group equation
was derived by Peliti and Liebler \cite{PL86}, Kleinert \cite{KL86} and
Polyakov \cite{PO86}. The explicit form of the one-loop RG  equation is
$$ \frac{d\kappa}{d\xi}=-\frac{3T}{4\pi} \,. $$
Here $\xi=\ln(r/a)$ and $r$ is the characteristic scale. As follows from the
equation the role of the dimensionless coupling constant is played by the
quantity
\begin{eqnarray}
g=\frac{3T}{4\pi\kappa}  \,.
\label{i1} \end{eqnarray}
Note that $\ln(\xi_p/a)\sim g^{-1}$. For real membranes
$g\sim10^{-2}-10^{-3}$ and consequently we can treat $g$ as a small
parameter. The smallness of $g$ means that there exists a wide range of
scales $r<\xi_p$ where thermal fluctuations can be treated in the framework
of the perturbation theory.

Corrugations of the membranes in a double film can be decomposed into
undulation (or bending) deformations and the squeezing ones. The bending
deformations are characterised by the displacement $u$ of the film as a whole
from its equilibrium position and the squeezing deformation is characterised
by variations of the film thickness $h$ (which is the separation between the
membares). We will believe that in equilibrium the film lies along the $X-Y$
plane. Then in the harmonic approximation it follows from (\ref{a1}) that the
energy of the film is
\begin{equation}
{\cal H}=\int dx\, dy\, \left[\kappa(\nabla^2u)^2+
\frac{\kappa}{4}(\nabla^2h)^2\right] \,,
\label{e1} \end{equation}
where both $u$ and $h$ are treated as functions of $x$ and $y$ and
$\nabla$ is the two-dimensional gradient.

At deriving (\ref{e1}) we neglected the interaction between the membranes.
First, one should remember about steric interaction associated with a certain
restriction of accessible configurations for one membrane in the presence of
the second membrane \cite{HE75}. The explicit expression for the energy is
\cite{HE78}
\begin{equation}
{\cal H}_{\rm ster}=\int dx\, dy\,
\frac{3\pi^2T^2}{128\kappa h^2} \,.
\label{eg2} \end{equation}
Due to the interaction (\ref{eg2}) two membranes can be treated as
independent only on scales smaller than $g^{-1/2}h$. Therefore (\ref{e1}) is
the main contribution to the energy if
\begin{equation}
qh>\sqrt g \,,
\label{e2} \end{equation}
where $q$ is the characteristic wave vector. Second, we should take into
account the Van der Waals interaction. We assume that the same liquid is
inside and outside the film. Then the Van der Waals energy is \cite{IZ85}
\begin{equation}
{\cal H}_{\rm vdw}=\int dx\,dy\,\frac{Ha^2}{2\pi h^4} \,,
\label{eg3} \end{equation}
where $H$ is the Hamaker constant. We can neglect the energy in comparison
with (\ref{e1}) if
$$ (qh)^4>\frac{H}{\kappa}\left(\frac{a}{h}\right)^2 \,. $$
In the following we will believe that the thickness of the film is large
enough for the inequality
$$ g^2>\frac{H}{\kappa}\left(\frac{a}{h}\right)^2 \,, $$
to be satisfied. Then (\ref{e2}) is the only restriction enabling us to
treat the energy (\ref{e1}) as the main contribution to the film energy.

\section{Dynamics}

We will examine the dynamics of the double membrane film in the
long-wavelength limit $qh\ll1$ where $q$ is the wave vector of the eigen
modes of the film. Note that the inequality $qh\ll1$ is compatible with
(\ref{e2}) since $g\ll1$. In the limit $qh\ll1$ one should take into account
the following variables describing the dynamics: the velocity of the film
${\bbox v}$, the displacement of the film $u$, the film thickness $h$ and the
densities of both membranes since they are conserved quantities. We will be
interested mainly in the squeezing mode associated with the relaxation of the
thickness $h$.

To find dynamical characteristics of the film one should solve the
conventional hydrodynamic equations in bulk supplemented by boundary
conditions on both membranes. In the linear approximation the problem was
solved by Brochard and Lennon \cite{BL75}, they found the dispersion law
of the squeezing mode
\begin{eqnarray}
\omega=-i\frac{\kappa h^3 q^6}{24\eta} \,,
\label{a11} \end{eqnarray}
where $\omega$ is the frequency of the mode and $\eta$ is the viscosity of
the liquid surrounding the membranes. At deriving (\ref{a11}) one assumed
that in equilibrium the film is flat. Note also the dispersion law
\cite{BL75}
\begin{eqnarray}
\omega=-i\frac{\kappa q^3}{2\eta} \,,
\label{e11} \end{eqnarray}
of the bending mode also found in the linear approximation. Note that the
dispersion law (\ref{a11}) is correct only if to neglect the direct
interaction of the membares that is at the condition (\ref{e2}) whereas
the region of applicability of the dispersion law (\ref{e11}) does not
depend on the interaction of the membranes since they move in-phase in
the bending mode. The elastic modes associated with variations of the
membrane densities are harder than (\ref{a11},\ref{e11}) \cite{KL93}.
Therefore the only effect of the elastic degrees of freedom at examining
the squeezing mode is the incompressibility condition
\begin{eqnarray}
\nabla_\alpha v_\alpha=0 \,.
\label{u11} \end{eqnarray}
Here and below we believe that all variables characterising the film are
functions of $x$ and $y$ and assume that Greek subscripts run over $x$ and
$y$.

We will consider the renormalization of the dispersion law (\ref{a11})
of the squeezing mode due to fluctuational effects. For the purpose nonlinear
dynamical equations of the film should be utilised. In the long-wavelength
limit $qh\ll1$ the equations can be derived phenomenologically. The reactive
(non-dissipative) part of the equations can be found using the Poisson
brackets method (see \cite{DV80} and also \cite{KL93}) whereas the
dissipative part of the equations is expressed via kinetic coefficients.
One should know the expression for the energy ${\cal H}$ of the system for
writing both contributions. Actually the expression only for one Poisson
bracket will be needed for us:
\begin{equation}
\left\{j_\alpha(x_1,y_1),h(x_2,y_2)\right\}=h(x_1,y_1)
\nabla_\alpha[\delta(x_1-x_2)\delta(y_1-y_2)] \,,
\label{a3} \end{equation}
where $j_\alpha$ is the two-dimensional momentum density of the film.
The expression (\ref{a3}) is characteristic of any scalar conserved quantity
of a film \cite{KL93}, the expression (\ref{a3}) is motivated by the fact
that the two-dimensional mass density of the film is $\rho h$ where $\rho$ is
the three-dimensional density of the liquid. Note that
$j_\alpha\approx\rho hv_\alpha$ since we believe that the membrane thickness
$a$ can be neglected in comparison with the film thickness $h$.

The dynamic equation for the thickness $h$ has the standard form following
from (\ref{a3})
\begin{eqnarray}
\partial_t h+\nabla_\alpha (v_\alpha h)=
\Gamma\nabla^2\frac{\delta{\cal H}}{\delta h} \,,
\label{a4} \end{eqnarray}
where $\partial_t\equiv\partial/\partial t$ and $\Gamma$ is the kinetic
coefficient. The second power of the gradient appeared in (\ref{a4}) since
the equation should support the conservation law of the liquid inside the
film and therefore the right-hand side of the equation should be a full
derivative at any ${\cal H}$. In the linear approximation we can neglect the
sweeping term in (\ref{a4}) due to (\ref{u11}). Substituting the harmonic
expression (\ref{e1}) for the energy ${\cal H}$ into (\ref{a4}) and comparing
the result with (\ref{a11}) one gets
\begin{equation}
\Gamma=h^3/(12\eta)\,.
\label{e3} \end{equation}
Note that $\Gamma$ is inversely proportional to the shear viscosity
coefficient. The point is that the dissipation described by $\Gamma$
comes from viscous motion of the liquid surrounding the double membrane
film which is excited hardly at large $\eta$.

The dynamic equation for $j_{\alpha}$ has the following form \cite{KL94}
\begin{equation}
\partial_tj_\alpha-\{{\cal H},j_\alpha\}=J_\alpha \,,
\label{i2} \end{equation}
where ${\bbox J}$ is the momentum flow from the bulk to the film. Just the
term supplies the main dissipation of the film momentum and therefore we
neglected the internal viscosity. In the linear approximation \cite{KL94}
\begin{equation}
J_\alpha=-2\eta\hat q v_\alpha\,,
\label{i3} \end{equation}
where $\hat q$ is the non-local operator which is reduced to multiplying by
the absolute value of the wave vector $q$ in the Fourier representation. The
Poisson bracket $\{{\cal H},j_\alpha\}$ can be reduced to the divergence of
the symmetric stress tensor for any energy ${\cal H}$ \cite{KL93}. Actually
only the contribution associated with the Poisson bracket (\ref{a3}) and
created by the harmonic energy (\ref{e1}) is relevant for us. Then the
equation (\ref{i2}) is written as
\begin{equation}
\partial_tj_\alpha+\frac{\kappa}{2}h\nabla_\alpha\nabla^4h
=-2\eta\hat q v_\alpha \,.
\label{a5} \end{equation}

We will not present here dynamical equations for the variables $j_z$, $u$ and
the densities of the membranes. The reason is that the equations for $j_z$
and $u$ describing the bending mode decouple in the approximation needed for
us from the equations (\ref{a4},\ref{a5}). Actually the equations for $j_z$
and $u$ are the same as for a single membrane, the corresponding nonlinear
equations can be found in \cite{KL93} and also in \cite{LM89,KL94}. As to
the equations for the densities of the membranes they need a separate
consideration which will be presented elsewhere. The only role of the degrees
of freedom at analysing the squeezing mode is reduced to the
incompressibility condition (\ref{u11}).

\section{Renormalization of squeezing mode}

As is seen from (\ref{a11}) in the long-wavelength limit the squeezing mode
is very soft. This is the reason why one anticipates that fluctuational
effects related to the mode are relevant. The effects are associated with
nonlinear terms in dynamic equations and can be examined in terms of the
diagrammatic technique of the type first developed by Wyld \cite{WY61} who
studied velocity fluctuations in a turbulent fluid. In the work \cite{MS73}
the Wyld technique was generalised for a broad class of dynamical systems. A
textbook description of the diagram technique can be found in the book by Ma
\cite{MA76}. The diagram technique can be formulated in terms of path
integrals as was first suggested by de Dominicis \cite{DO76} and Janssen
\cite{JA76}. In the framework of this approach apart from conventional
dynamic variables one should introduce also auxiliary fields conjugated to
the variables. Then dynamical correlation functions of the variables  can be
presented as functional integrals over both type of fields: conventional and
auxiliary. The integrals are taken with the weight $\exp(i{\cal I})$, where
${\cal I}$ is an effective action which is constructed on the basis of
nonlinear dynamic equations of the system.

Being interested in the renormalization of the squeezing mode of the double
membrane film we will take into account only the variables $h$ and $v_\alpha$
and the corresponding auxiliary conjugated fields $p$ and $\mu_\alpha$. We
should also remember about the incompressibility condition (\ref{u11}) and
impose the analogous constraint $\nabla_\alpha\mu_\alpha=0$ on the field
$\mu_\alpha$. Then, say, the correlation function of the film thickness $h$
is written as
\begin{eqnarray}
\langle h_1 h_2 \rangle =
\int{\cal D}h\,{\cal D}{\bbox v}_{\rm tr}
{\cal D}p\,{\cal D}{\bbox\mu}_{\rm tr}
\exp\left(i {\cal I}\right) h_1 h_2 \,,
\label{b3} \end{eqnarray}
where the subscript `tr' implies that in the Fourier representation we
should take only transverse to the wave vector ${\bbox q}$ components of the
fields ${\bbox v}$ and ${\bbox\mu}$. The explicit expression for the
effective action figuring in (\ref{b3}) can be found using the dynamical
equations (\ref{a4},\ref{a5}). It can be written as the sum of the reactive
and the dissipative parts ${\cal I}={\cal I}_{\rm reac}+{\cal I}_{\rm diss}$
where
\begin{eqnarray} &&
{\cal I}_{\rm reac}=\int dt\,
d^2r\, \left\{p\partial_th+pv_\alpha\nabla_\alpha h
+\mu_\alpha\partial_tj_\alpha
-\frac{\kappa}{2}\mu_\alpha\nabla^4h\nabla_\alpha h\right\} \,,
\label{b1} \\ &&
{\cal I}_{\rm diss}=\int dt\, d^2r\,
\left\{-\frac{1}{2}\Gamma{\kappa}p\nabla^6h
+iT\Gamma(\nabla p)^2
+2\eta{\bbox\mu}\hat q({\bbox v}
+iT{\bbox\mu})\right\} \,.
\label{b2} \end{eqnarray}
The detailed derivation of the effective action for the problem can be found
in \cite{LM89,KL94}.

Let us introduce the designations for the pair correlation functions. Taking
into account only the transverse components of the fields ${\bbox v}$ and
${\bbox\mu}$ we can write
\begin{eqnarray} &&
\langle h(t,{\bbox r})p(0,{\bbox 0})\rangle
=\int\frac{d\omega\,d^2q}{(2\pi)^3}
\exp(-i\omega t+i{\bbox q}{\bbox r})G(\omega,{\bbox q}) \,,
\nonumber \\ &&
\left\langle v_\alpha(t,{\bbox r})
\mu_\beta(0,{\bbox 0})\right\rangle
=\int\frac{d\omega\,d^2q}{(2\pi)^3}
\exp(-i\omega t+i{\bbox q}{\bbox r})
\left[\delta_{\alpha\beta}-\frac{q_\alpha q_\beta}{q^2}\right]
G_{\rm tr}(\omega,{\bbox q}) \,,
\label{ma6} \\ &&
\langle h(t,{\bbox r})h(0,{\bbox 0})\rangle
=\int\frac{d\omega\,d^2q}{(2\pi)^3}
\exp(-i\omega t+i{\bbox q}{\bbox r})D(\omega,{\bbox q}) \,,
\nonumber \\ &&
\left\langle v_\alpha(t,{\bbox r})
v_\beta(0,{\bbox 0})\right\rangle
=\int\frac{d\omega\,d^2q}{(2\pi)^3}
\exp(-i\omega t+i{\bbox q}{\bbox r})
\left[\delta_{\alpha\beta}-\frac{q_\alpha q_\beta}{q^2}\right]
D_{\rm tr}(\omega,{\bbox q}) \,.
\label{ma7} \end{eqnarray}
The correlation functions $\langle pp\rangle$ and $\langle\mu\mu\rangle$ are
equal to zero (what is the general property of the technique, see e.g.
\cite{KL93}). The functions $D$ and $D_{\rm tr}$ determine the pair
correlation functions of the observable quantities and the functions $G$,
$G_{\rm tr}$ are response functions. Therefore, say, the function $G(\omega)$
is analytical in the upper $\omega$ half plane.

It is possible to formulate the diagram technique for calculating correlation
functions (\ref{ma6},\ref{ma7}). The harmonic part of the effective action
${\cal I}={\cal I}_{\rm reac}+{\cal I}_{\rm diss}$ determines the `bare'
values of the response functions
\begin{equation}
G_0(\omega,{\bbox q})
=-\frac{1}{\omega+i\Gamma\kappa q^6/2} \,, \qquad
G_{\rm tr,0}(\omega,{\bbox q})
=-\frac{1}{\rho h\omega+2i\eta q} \,.
\label{mm1} \end{equation}
The values of the `bare' pair correlation functions satisfy the relations
\begin{equation}
{\sl Im}\,G=\frac{\kappa q^4}{4T}D \,, \qquad
{\sl Im}\,G_{\rm tr}=\frac{1}{2T}D_{\rm tr} \,,
\label{ma10} \end{equation}
which are consequences of the fluctuation-dissipation theorem. Besides the
harmonic part the effective action ${\cal I}$ contains terms of the third
order which determine the third-order vertices which figure on diagrams
representing the perturbation series for the correlation functions
(\ref{ma6},\ref{ma7}). One can check the relations (\ref{ma10}) order by
order and consequently they are true for the `dressed' correlation functions
(\ref{ma6},\ref{ma7}). Note that
\begin{equation}
\int\frac{d\omega}{2\pi} D(\omega,q)
=\frac{2T}{\kappa q^4} \,.
\label{mm2} \end{equation}
The relation (\ref{mm2}) is a consequence of (\ref{ma10}), analyticity of
$G(\omega)$ in the upper half-plane and of the asymptotic law
$G(\omega)\approx-\omega^{-1}$ which is correct at large $\omega$.
Note that (\ref{mm2}) could be obtained directly from (\ref{e1}) since the
integral over frequencies is just the simultaneous correlation function.

The analysis of the diagrams shows that they contain infrared logarithms
related to the lines representing the correlation function $D$ (\ref{ma7}).
The lines produce factors
\begin{equation}
\langle\nabla_\alpha h(t, {\bbox r})
\nabla_\beta h(t, {\bbox 0}) \rangle
=\frac{TL}{2\pi\kappa}\delta_{\alpha\beta} \,,
\label{mm3} \end{equation}
where $L=\ln[hg^{-1/2}/r]$ and $r^{-1}$ is determined by characteristic
external wave vector of the diagram. The expression (\ref{mm3})
can be found from (\ref{mm2}) if to remember about the condition
(\ref{e2}). The presence of the logarithmic contributions implies
that the main renormalization of a correlation function like
$G(\omega,{\bbox q})$ is produced by the degrees of freedom with
the wave vectors much smaller than $q$. Therefore we should extract from the
diagrammatic expressions for $G(\omega,{\bbox q})$ only the contributions
corresponding to the interaction with the degrees of freedom.

The program can be realized directly on the language of the functional
integral. Let us separate the variables $h$, $p$, ${\bbox v}$, ${\bbox\mu}$
into fast parts (with wave vectors larger than $q$), basic parts (with the
wave vectors of the order of $q$) and slow parts (with wave vectors smaller
than $q$). At calculating $G(\omega,{\bbox q})$ we can forget about the fast
parts and keep the interaction of the basic part with the slow part. Then we
obtain from (\ref{b1},\ref{b2})
\begin{equation}
{\cal I}=\int dt\,d^2r\,
\left\{p\partial_th+pv_\alpha m_\alpha
+\mu_\alpha\partial_tj_\alpha
-\frac{\kappa}{2}\mu_\alpha\nabla^4hm_\alpha
-\Gamma\frac{\kappa}{2}p\nabla^6h
+2\eta{\bbox\mu}\hat q{\bbox v}\right\}+\dots \,,
\label{mm4} \end{equation}
where $h$, $p$, ${\bbox v}$, ${\bbox\mu}$ denote the basic parts of the
fields, $m_\alpha$ is the gradient of the slow part of $h$ and dots
designate irrelevant terms. The action (\ref{mm4}) is of the second order
over $h$, $p$, ${\bbox v}$, ${\bbox\mu}$ and consequently the integrals over
the fields can be taken explicitly. Since ${\bbox m}$ varies only weakly on
the length $q^{-1}$ we get
\begin{eqnarray} &&
G(\omega,{\bbox q})=-\left\langle
(\rho h\omega+2i\eta q)\Delta^{-1}\right\rangle_m \,,
\label{mm5} \\ &&
G_{\rm tr}(\omega,{\bbox q})=-\left\langle
(\omega+i\kappa\Gamma q^6/2)\Delta^{-1}\right\rangle_m \,,
\label{mm6} \\ &&
\Delta=(\rho h\omega+2i\eta q)(\omega+i\kappa\Gamma q^6/2)
-\kappa q^4 m_{\rm tr}^2/2 \,,
\label{mm7} \end{eqnarray}
where
$$ m_{\rm tr}^2=\left(\delta_{\alpha\beta}-
\frac{q_\alpha q_\beta}{q^2}\right)m_\alpha m_\beta \,, $$
and the designation $\langle\dots\rangle_m$ means averaging over statistics
of ${\bbox m}$. At calculating (\ref{mm5},\ref{mm6}) we substituted
${\bbox j}=\rho h{\bbox v}$. Actually the terms with $\rho h$ can be
neglected and we omit the terms below.

At averaging in (\ref{mm5},\ref{mm6}) the statistics of ${\bbox m}$ can be
regarded to be Gaussian. The point is that only simultaneous correlation
functions of ${\bbox m}$ enter the expressions and the functions are
described by the harmonic function (\ref{e1}). The pair correlation function
of ${\bbox m}$ is equal to (\ref{mm3}). Therefore
$$ \langle m_{\rm tr}^2 \rangle=\frac{TL}{2\pi\kappa}\,, $$
and we find from (\ref{mm5})
\begin{equation}
G(\omega,{\bbox q})=-\int\limits_{-\infty}^{+\infty}
\frac{d\zeta}{\sqrt{2\pi}}\exp\left(-\zeta^2/2\right)
\left(\omega+i\frac{\kappa\Gamma}{2}q^6
+i\frac{TL}{8\pi\eta}q^3\zeta^2\right)^{-1} \,.
\label{mm8} \end{equation}
We see that $G$ as a function of the frequency $\omega$ have the cut along
the imaginary semi-axis which starts from $\omega=-i\Gamma\kappa q^6/2$ and
goes up to $-i\infty$. The effective length of the cut can be estimated as
$Tq^3/\eta$ what is the new characteristic frequency related to fluctuations.
Let us compare the frequency with the position of the pole in the bare
expression
\begin{equation}
\frac{Tq^3/\eta}{\Gamma\kappa q^6}\sim\frac{g}{(qh)^3} \,.
\label{i6} \end{equation}
We conclude that fluctuation effects dominate in the region
$g^{1/2}<qh<g^{1/3}$. Now, we can justify neglecting $\rho h\omega$ in
comparison with $\eta q$ in the above expressions. At $qh\sim1$
$$ \rho h\omega/(\eta q)\sim \rho\kappa/(\eta^2 h)\sim a/h\ll1 \,, $$
and at $qh\sim\sqrt g$
$$ \rho h\omega/(\eta q)\sim\rho\kappa/(\eta^2 h)g^2\ll1 \,. $$

Performing Fourier transform of (\ref{mm8}) over frequencies one gets
\begin{equation}
G(t,{\bbox q})=i\left(1+\frac{TL}{4\pi\eta}q^3t\right)^{-1/2}
\exp\left\{-\frac{\kappa}{2}\Gamma q^6t\right\} \,.
\label{i7} \end{equation}
The expression (\ref{i7}) is correct for positive time $t$, for negative
times $G(t)=0$ due to the causality principle since $G$ is the response
function. We see from (\ref{i7}) that in the fluctuation region
$g^{1/2}<qh<g^{1/3}$ there appears an intermediate power asymptotics
$t^{-1/2}$ which at large times $t$ is changed by the exponential decay.
That means that the squeezing mode is described by a nonlocal in time
dynamic equation.

The above assertion is correct for the wave vectors $q\gtrsim\sqrt{g}\,/h$. In
the limit $qh\ll\sqrt{g}$ we return to the local equation (\ref{a4}) but with
the renormalised kinetic coefficient $\tilde\Gamma$. The quantity can be
found if to integrate the weight $\exp(i{\cal I})$ over the degrees of
freedom with the wave vectors $q\gtrsim\sqrt{g}\,/h$. The main effect
is related to the sweeping term in the effective action (\ref{b1}).
Due to the integration over the degrees of freedom with the
wave vectors $q\gtrsim\sqrt g\,/h$ the term $iT\Gamma(\nabla p)^2$ in
(\ref{b2}) for the long-wavelength degrees of freedom is renormalised and
we found for the renormalised value of $\Gamma$:
\begin{equation}
\tilde\Gamma-\Gamma=\frac{1}{4T}\int dt\, d^2 r\,
\langle{\bbox v}(t,{\bbox r})h(t,{\bbox r})
{\bbox v}(0,{\bbox 0})h(0,{\bbox 0})\rangle \,,
\label{ren} \end{equation}
where averaging is performed over the degrees of freedom with the wave
vectors $q\gtrsim\sqrt g\,/h$. Using the renormalised expressions for the
correlation functions we get the estimate
$\tilde\Gamma\sim g^{-1/2}\Gamma\gg\Gamma$.

\section{Conclusion}

We demonstrated that fluctuations essentially modify the character of the
squeezing mode due to its nonlinear coupling with transversal shear
hydrodynamic mode. Fluctuation effects lead to non locality of the equation
for the mode, the corresponding Green function is (\ref{i7}). The new
characteristic frequency of the mode related to fluctuations is
$\omega\sim Tq^3/\eta$ ($q$ is the wave vector) which remarkably does not
depend on bending elasticity. It is important to distinguish the
characteristic frequency from the attenuation of the membrane bending mode
(\ref{e11}) having the same $q^3$ dependence on the wave vector. Let us
stress that the strong fluctuation effects are observed only for dynamics.
Static characteristics are not influenced by fluctuations because of the
smallness of the coupling constant (\ref{i1}). That is the reason why only
the harmonic part of the energy (\ref{e1}) is needed for us.

Strong dynamic fluctuations of $h$ occur for the wave vectors
$q\gtrsim\sqrt g\,/h$. For smaller wave vectors fluctuations of $h$ are
weak. Nevertheless even for the wave vectors there is a memory of the region
of strong fluctuations which is the renormalised value of the kinetic
coefficient $\Gamma$ in the equation (\ref{a4}): The `bare' value (\ref{e3})
is substituted by $\tilde\Gamma\sim g^{-1/2}\Gamma\gg\Gamma$. Note also
that to analyse the dispersion law of the squeezing mode in the limit
$q\ll\sqrt g\,/h$ starting from (\ref{a4}) one should take into account
besides the energy (\ref{e2}) also the steric (\ref{eg2}) and the
Van-der-Waals (\ref{eg3}) contributions to the energy. As a result one finds
$$ \omega=-i\tilde\Gamma q^2 \left(\frac{9\pi^2T^2}{64\kappa h^4}
+\frac{10 H a^2}{\pi h^6}\right) \,. $$

Let us discuss possibilities to check our predictions experimentally.
The membranes can be studied by a variety of experimental techniques.
Lately laser ``tweezers" are becoming a useful tool for probing dynamical
properties of membranes. The technique enables one to obtain direct
information about amplitudes and characteristic times of dynamical
fluctuations for different objects constituted from membranes. For details
see the monography \cite{BL90} and recent experiments \cite{BM94,BM95,BFM95}.
We can mention as well force apparatus measurements \cite{RM96} making
possible to investigate dynamical response for two very thin lamellar systems
confined between the walls and the classical light scattering experiments.
Because of relaxation of membrane fluctuations, the scattered light has a
broadened spectral distribution compared with the incident light. Despite the
broadening is small, the modern technique of light beating (intensity
fluctuation spec\-tro\-sco\-py) is allowed to obtain information about
eigen-modes of the system.

The conclusions concerning the renormalization of the squeezing mode in our
opinion are interesting both in its own right and as a new test of the
membrane fluctuations.

\acknowledgements

The research described in this publication was made possible in
part by RFFR Grants. E.K. thanks Max-Planck Institute for Physics
of Complex Systems (Dresden) for supporting his stay at this institute.

\end{document}